\documentclass[prl,aps,twocolumn,showpacs]{revtex4}

\usepackage{epsfig}

\begin{document}

\title{Microscopic coexistence of superconductivity and antiferromagnetism 
in underdoped Ba(Fe$_{1-x}$Ru$_{x}$)$_2$As$_2$}

\author{Long Ma$^{1}$}
\author{G. F. Ji$^{1}$}
\author{Jia Dai$^{1}$}
\author{X. R. Lu$^{1}$}
\author{M. J. Eom$^{2}$}
\author{J. S. Kim$^{2}$}
\author{B. Normand$^{1}$}
\author{Weiqiang Yu$^{1}$}
\email{wqyu_phy@ruc.edu.cn}
\affiliation{$^{1}$Department of Physics, Renmin University of China, Beijing 
100872, China\\
$^{2}$Department of Physics, Pohang University of Science and Technology, 
Pohang 790-784, Korea}

\date{\today}


\begin{abstract}

We use $^{75}$As nuclear magnetic resonance (NMR) to investigate the local 
electronic properties of Ba(Fe$_{1-x}$Ru$_{x}$)$_2$As$_2$ ($x =$ 
0.23). We find two phase transitions, to antiferromagnetism at $T_N \approx$ 
60 K and to superconductivity at $T_C \approx$ 15 K. Below $T_N$, our data 
show that the system is fully magnetic, with a commensurate antiferromagnetic 
structure and a moment of 0.4 $\mu_B$/Fe. The spin-lattice relaxation rate 
$1/^{75}T_1$ is large in the magnetic state, indicating a high density of 
itinerant electrons induced by Ru doping. On cooling below $T_C$, $1/^{75}T_1$ 
on the magnetic sites falls sharply, providing unambiguous evidence for 
the microscopic coexistence of antiferromagnetism and superconductivity. 

\end{abstract}

\maketitle

In the iron-based superconductors, superconductivity 
(SC) is achieved on suppressing a long-ranged antiferromagnetic order 
\cite{Dai_Nature_453_899} by doping or pressure. At this phase boundary, 
much attention has been drawn to the question of whether SC may coexist 
with antiferromagnetism (AFM). Proposals for possible coexisting phases 
have included commensurate \cite{Wiesenmayer_PRL_107} and incommensurate 
\cite{Pratt_coexist, Laplace_PRB_80, Marsik_PRL_105} magnetic structures, 
competition between AFM and SC \cite{Wiesenmayer_PRL_107, Laplace_PRB_80, 
Marsik_PRL_105, Nandi_PRL_104, Iye_JPSJ}, and variations in the size of 
the ordered moment \cite{Drew_NM_8, Bao_11020830} or the pairing symmetry 
\cite{Fernandes_PRB_82_014521, Maiti_arxiv_1203_0991}. No consensus has 
yet been reached on the pairing mechanism or the possible phenomena arising 
from the interplay of AFM and SC. For most materials, local-probe studies 
on high quality samples are required as a matter of urgency to distinguish 
the key properties of microscopic coexistence from any form of phase 
separation.

Because Ru is a $4d$ counterpart of Fe, it is expected to be an isovalent 
substituent. Isovalent substitution of P for As has been used to investigate 
the transition between \cite{Nakai_PRL}, and the possible coexistence of 
\cite{Iye_JPSJ}, AFM and SC in BaFe$_2$(As$_{1-x}$P$_x$)$_2$ powders. 
The suppression of AFM by doping appears to be rather slow also in 
Ba(Fe$_{1-x}$Ru$_x$)$_2$As$_2$, and both transport \cite{Thaler_PRB_82_014534} 
and angle-resolved photoemission (ARPES) studies \cite{Dhaka_prl, 
Brouet_PRL_105_087001, Xu_12034699} indicate that changes in the band 
structure with Ru doping are small. Crucially, some signs of coexistence 
between AFM and SC have been observed by bulk probes \cite{Kim_PRB_83_054514, 
Kim_JS} in the underdoped region $0.15 \le x \le 0.3$. This slow suppression 
of AFM and the broad regime of coexistence offer a valuable opportunity to 
seek out and characterize a microscopic coexistence with minimal effects 
from doping inhomogeneity. 

In this letter, we use $^{75}$As NMR as a local probe to investigate Ru 
doping effects and the microscopic coexistence of AFM and SC in single 
crystals of Ba(Fe$_{0.77}$Ru$_{0.23}$)$_2$As$_2$. This underdoped system has 
several properties quite distinct from the parent compound BaFe$_2$As$_2$. 
Although the doping produces no strong changes in carrier density and 
magnetic correlations above $T_N$, below this the $^{75}$As spectrum shows 
fully magnetic signals and the AFM structure is commensurate, with a large 
moment of approximately 0.4 $\mu_B$/Fe. Below $T_C$, the spin-lattice 
relaxation rate $1/^{75}T_1$ shows a superconducting gap opening on the 
magnetic sites, providing incontrovertible local evidence for the microscopic 
coexistence of AFM and SC. We observe a high density of itinerant electrons 
below $T_N$, which constitutes clear experimental evidence for the effect of 
Ru doping in the magnetic state and offers an important clue to understanding 
the interplay of SC and magnetic order in Fe-based superconductors.  

We begin by considering the nature of the evidence for coexistence. 
Much of the discussion in pnictides concerns non-isovalent doping of 
BaFe$_2$As$_2$, particularly Ba(Fe$_{1-x}$Co$_x$)$_2$As$_2$ \cite{rcb}, 
where the reported coexistence occurs over a very narrow range of doping. 
In most experiments, the evidence is indirect, in that $\mu$SR is used to 
probe magnetic order but bulk measurements are used for the superconductivity;
NMR studies showing bulk superconductivity on the magnetic sites, with no 
mixing of paramagnetic contributions, is still lacking \cite{Laplace_PRB_80, 
Julien_EPL_87_37001}. The situation is complicated by the question of sample 
homogeneity \cite{Wiesenmayer_PRL_107, Julien_EPL_87_37001, Bao_11020830}, 
an extreme example of which has emerged recently in the ``microscopic phase 
separation'' of alkali-intercalated FeSe \cite{ramzea}. Here we provide direct 
and unambiguous evidence for the coexistence of AFM and SC by using NMR, a 
local probe sensitive to both types of order on the atomic scale. In the 
process we will eliminate any sample inhomogeneity-induced phase separation 
of AFM and SC regions, and also any type of concomitant proximity-effect SC 
in AFM regions. 

Returning to our target material, Ba(Fe$_{1-x}$Ru$_x$)$_2$As$_2$, the 
isovalent suppression of AFM by Ru doping is so far not well understood. 
Theoretically, it has been proposed that the more extended $d$-orbitals 
on the Ru sites should weaken AFM order \cite{Zhang_PRB_79_174530}. 
Experimental investigation is required to address whether this weakening 
is caused by a chemical pressure \cite{Thaler_PRB_82_014534}, a magnetic 
dilution \cite{Brouet_PRL_105_087001}, an effective doping, or a combination 
of effects, and the properties of the phases may vary strongly under the 
different scenarios. As one example, a nodal superconductivity has been 
suggested in underdoped Ba(Fe$_{1-x}$Ru$_x$)$_2$As$_2$ \cite{Qiu_PRX_2_011010}, 
and controversy remains over its possible origin in the reduced $c$-axis 
lattice parameter \cite{Qiu_PRX_2_011010}, in a magnetic effect 
\cite{Maiti_arxiv_1203_0991}, or in some other coexistence phenomenon. 

Our Ba(Fe$_{0.77}$Ru$_{0.23}$)$_2$As$_2$ single crystals were synthesized 
by the flux-grown method with FeAs as flux \cite{Kim_JS}. We stress here 
and below that bulk (near 100\%) superconductivity has been reported in 
crystals grown in the same batch as ours, on the basis of thermal 
conductivity measurements \cite{Qiu_PRX_2_011010}. The chemical 
composition was determined by energy-dispersive x-ray (EDX) measurements. 
For our NMR experiments, single crystals with dimensions of approximately 
3$\times$1.5$\times$0.1 mm$^3$ were chosen. The NMR measurements were 
performed on $^{75}$As nuclei ($S =$ 3/2) under a magnetic field of 10 T 
and in two different field orientations. The frequency-swept NMR spectra 
were obtained by integrating the Fourier transform of the spin-echo 
intensity. The spin-lattice relaxation rate $1/^{75}T_1$ is measured 
with the inversion-recovery method, by fitting the magnetization to 
the functional form $I(t)/I(0) = 1 - a (0.1e^{-t/T_{1}} + 0.9e^{-6t/T_{1}})$.

\begin{figure}
\includegraphics[width=7cm, height=7cm]{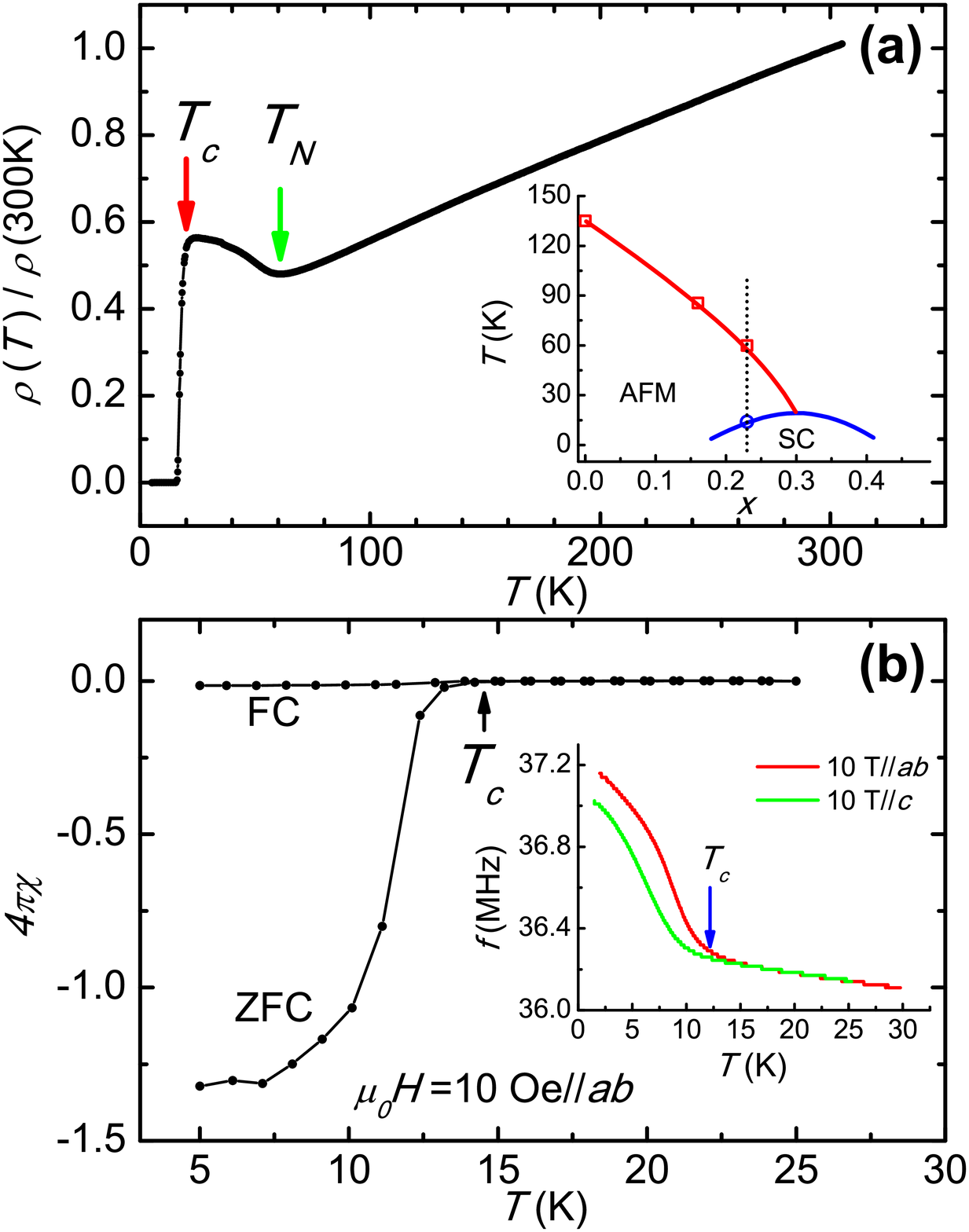}
\caption{\label{suscep1}(color online) (a) In-plane resistivity of the 
Ba(Fe$_{0.77}$Ru$_{0.23}$)$_2$As$_2$ crystal on cooling at zero field. Inset: 
schematic phase diagram of Ba(Fe$_{1-x}$Ru$_{x}$)$_2$As$_2$, where the dotted 
line represents the doping of our crystals. (b) d.c.~susceptibility during 
field-cooling (FC, with a 10 Oe field applied in the crystalline $ab$-plane) 
and zero-field-cooling (ZFC). Inset: resonance frequency of the NMR circuit 
as a function of temperature, with a 10 T field applied in two orientations.}
\end{figure}
   
Our Ba(Fe$_{0.77}$Ru$_{0.23}$)$_2$As$_2$ sample is clearly underdoped, as 
shown by the presence of two separate transitions in the resistivity and 
the susceptibility, presented respectively in Figs.~\ref{suscep1}(a) 
and~\ref{suscep1}(b). The in-plane resistivity $\rho(T)$ decreases almost 
linearly with temperature from 300 K down to 60 K. The low-temperature 
upturn in $\rho(T)$ indicates the simultaneous structural and AFM transition 
at $T_N \approx$ 60 K \cite{Kim_PRB_83_054514, Kim_JS}, and the sharp drop 
of $\rho(T)$ to zero is the superconducting transition at $T_C \approx$ 
15 K. A schematic phase diagram for Ba(Fe$_{1-x}$Ru$_{x}$)$_2$As$_2$, based 
on the data of Ref.~\cite{Kim_PRB_83_054514}, is shown in the inset of 
Fig.~\ref{suscep1}(a); the squares are our own data (not shown) for samples 
of different dopings, and the sample used for our NMR experiments is clearly 
in the underdoped regime. The d.c.~susceptibility shows full demagnetization 
just below $T_C$ in Fig.~\ref{suscep1}(b). Bulk superconductivity under the 
10 T NMR field is also monitored {\it in situ} by the resonance frequency of 
the NMR circuit, as shown in the inset of Fig.~\ref{suscep1}(b). The rapid 
increase in frequency below $T_c$, which is reduced to 12 K at 10 T, 
represents a decrease in the a.c.~susceptibility due to superconductivity. 

\begin{figure}
\includegraphics[width=9cm, height=6cm]{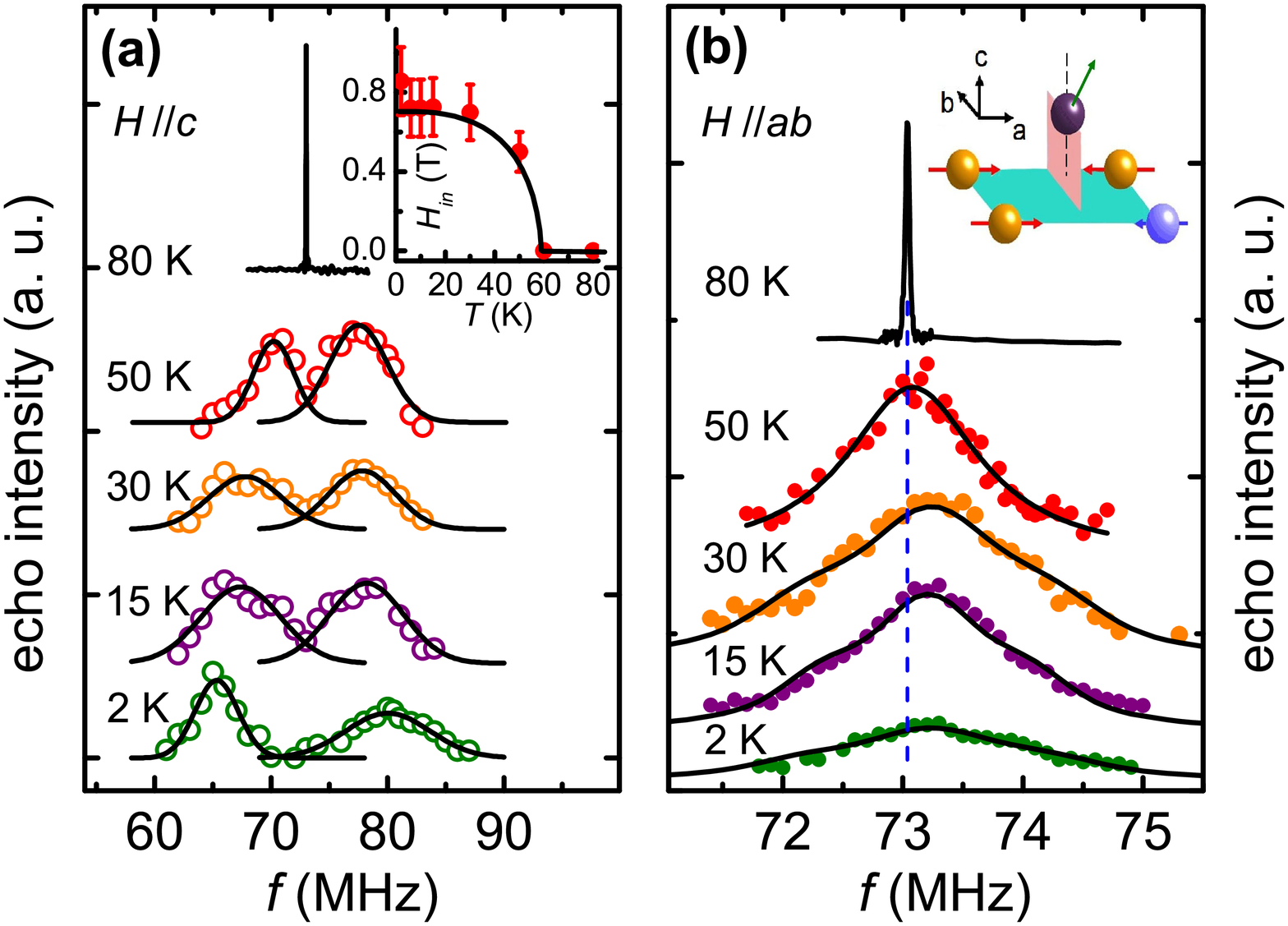}
\caption{\label{spec2}(color online) $^{75}$As NMR spectra with a 10 T 
magnetic field applied (a) along the $c$-axis and (b) in the $ab$-plane of our 
single crystal. Inset panel (a): static hyperfine field at $^{75}$As sites as 
a function of temperature. Inset panel (b): representation of one magnetic 
configuration in the presence of Ru dopants and the resulting $^{75}$As 
hyperfine field.}  
\end{figure}

The $^{75}$As spectra for a range of temperatures are shown in both field 
orientations in Figs.~\ref{spec2}(a) and~\ref{spec2}(b). For $H \parallel c$, 
the full width at half maximum (FWHM) height of the central transition is 
approximately 40 kHz at $T =$ 80 K. The spectrum broadens significantly and 
splits into two separate peaks on cooling below $T_N =$ 60 K, as expected at 
the AFM transition. For $H \parallel ab$, the FWHM is 42 kHz at $T =$ 80 K, 
and also broadens below $T_N$, the single peak shifting to a slightly higher 
frequency. 

The spectral splitting we observe below $T_N$ for $H \parallel c$ and 
the frequency shift for $H \parallel ab$ both demonstrate that our 
crystal is fully magnetic, with a commensurate A-type (striped) AFM order. 
Located above the centers of the Fe squares, the $^{75}$As nuclei detect an 
off-diagonal contribution from Fe moments in the stripe configuration, and 
the internal static hyperfine field is oriented along the $c$-axis 
\cite{Kita_JPSJ_77_114709}. The $^{75}$As spectrum below $T_N$ 
is therefore split, with two peak frequencies $f = \gamma H_{total} = \gamma 
(H_{ext} \pm H_{in})$, for $H \parallel c$, whereas for $H \parallel ab$ there 
is a spectral shift with one peak frequency $f = \gamma H_{total} = \gamma 
\sqrt{H^2_{ext} + H^2_{in}}$; $H_{ext}$ and $H_{in}$ refer respectively to the 
applied field and the internal hyperfine field. For $H \parallel c$, the 
spectral weight at the paramagnetic resonance frequency decreases rapidly 
below $T_N$, essentially vanishing below 30 K [Fig.~\ref{spec2}(a)]. This 
observation means that every ion in the sample is magnetic. It also excludes 
incommensurate magnetism, as suggested \cite{Kitagawa_JPSJ} in NaFeAs, which 
would also require a much higher spectral weight at this frequency. Similarly 
for $H \parallel ab$, the spectrum at this frequency [dashed line in 
Fig.~\ref{spec2}(b)] contains no detectable residual peak below $T_N$. 

We estimate the internal field $H_{in}$ from the shift in peak frequencies. 
As shown in the inset of Fig.~\ref{spec2}(a), $H_{in}$ increases with 
decreasing temperature, reaching $H_{in} =$ 0.7 T at $T =$ 2 K. Under the 
assumption that the off-diagonal hyperfine coupling is the same as in 
BaFe$_2$As$_2$, $A_{hf}^{ac} =$ 0.43T/$\mu_B$ \cite{Kita_JPSJ_77_114709}, 
we estimate the magnetic moment as $H_{in}/4A_{hf}^{ac} \approx 0.4 \mu_B$/Fe 
at $T =$ 2 K, consistent with the neutron scattering result at a similar 
doping \cite{Kim_PRB_83_054514}. Further, the large line width for $H 
\parallel ab$ can be fully explained as a Ru doping effect: with one Ru 
atom in a square unit [inset of Fig.~\ref{spec2}(b)], the $^{75}$As nuclei 
pick up both $ab$-plane and $c$-axis field components due to the unbalanced 
hyperfine fields from the Ru and Fe moments. The solid lines in Fig.~2 were 
drawn by taking the statistical distribution of square units corresponding 
to 23\% Ru doping and applying a Gaussian broadening. If we assume further 
that $^{75}$As has the same hyperfine coupling to Ru as to Fe, we may estimate 
a lower bound for the hyperfine field as 0.25$\mu_B$/Ru at $T =$ 2 K. This 
result once again supports a fully magnetic state, but with a magnetic 
dilution effect on the Ru sites.

At this point we discuss in detail the question of doping inhomogeneity. 
This is a crucial issue in Fe superconductors, where the unexpected 
electronic properties of non-stoichiometric materials turn out in several 
cases to be due to inclusions of phases which differ in chemical composition. 
The most extreme example is alkali-intercalated FeSe, where the AFM and SC 
phases show a microscale phase separation \cite{ramzea}, and this is clearly 
a possibility we must eliminate in Ba(Fe$_{1-x}$Ru$_{x}$)$_2$As$_2$. In our data 
(Fig.~2), the $H \parallel c$ spectrum at 15 K shows some small but complicated
features beyond the statistical fit, which are probably caused by minor doping 
inhomogeneities; related strain or disorder effects are to be expected at 
this near-critical doping, where we believe the system is close to the 
transition to zero magnetic order. Above $T_N$, we find no decrease in 
spectral weight in the paramagnetic frequency range between 60 K and 120 K 
(data not shown), which indicates that the impact of inhomogeneity on the 
magnetism is weak. In general, we expect that the local Ru doping is related 
directly to the on-site moment, and hence to the width of the NMR lines below 
$T_N$. From our spectral-weight analysis, we estimate that the fraction of 
sites in a paramagnetic phase (zero internal field) is below 0.05\%, while 
for those with moments 0.7$-$0.8$\mu_B$/Fe (close to the undoped value) it 
is approximately 3\%. We stress that our data (Fig.~2) show no evidence for 
bimodal distributions, or any other clustering suggesting a favored doping 
value away from the average, which would be an essential sign of phase 
separation. We remind the reader that microscopic and bulk phase separation 
are essentially the same to a probe as local as NMR. We define microscopic 
coexistence as the statement that the majority of the sample hosts both SC 
and AFM order. It is therefore not excluded if small fractions at extreme 
doping inhomogeneities lack one type of order or the other. 

\begin{figure}
\includegraphics[width=8.5cm,height=5.3cm]{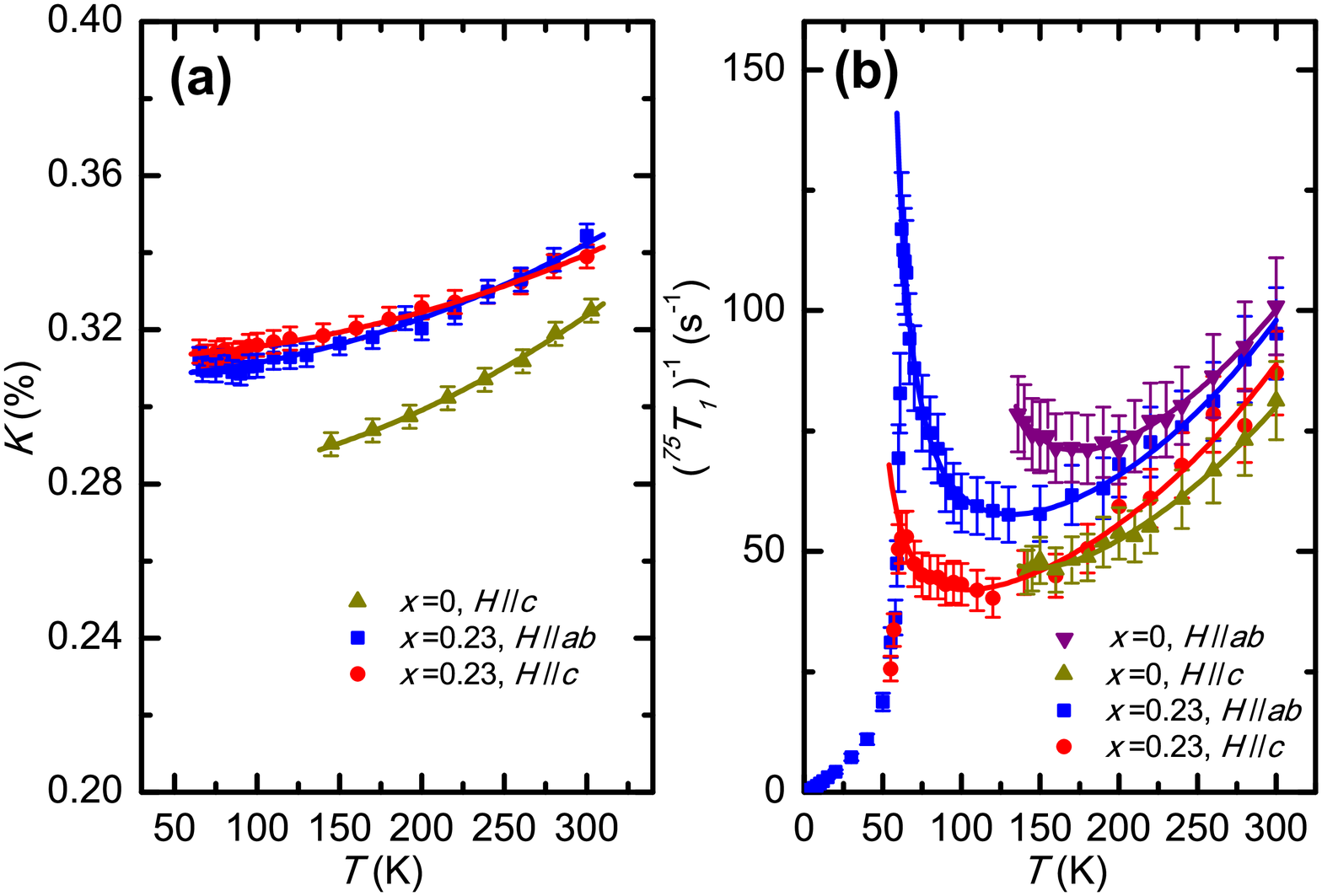}
\caption{\label{slrr3}(color online) (a) $^{75}K(T)$ as a function of 
temperature for Ba(Fe$_{1-x}$Ru$_{x}$)$_2$As$_2$ ($x =$ 0, 0.23) with two 
field orientations. The solid lines are a fit to the function $^{75}K(T)
 = a + bT^2$. (b) $1/^{75}T_1$ as a function of temperature with two field 
orientations. For each doping, the solid line above $T_N$ is a fit to 
$1/^{75}T_1 = AT/(T - \theta) + BT$.}  
\end{figure}

The fact that our NMR signals are fully magnetic below $T_N$ is already an 
indication for the coexistence of AFM and SC, but we defer a full discussion 
until our presentation of the spin dynamics (below). The Knight shift, 
$^{75}K(T)$, and spin-lattice relaxation rate, $1/^{75}T_1(T)$, at the peak 
frequency are shown in Figs.~\ref{slrr3}(a) and~\ref{slrr3}(b) for the 
two field orientations. We have also measured the same quantities for 
BaFe$_2$As$_2$ single crystals, and show the data in Fig.~\ref{slrr3} for 
comparison.

\begin{figure}
\includegraphics[width=8.5cm, height=9cm]{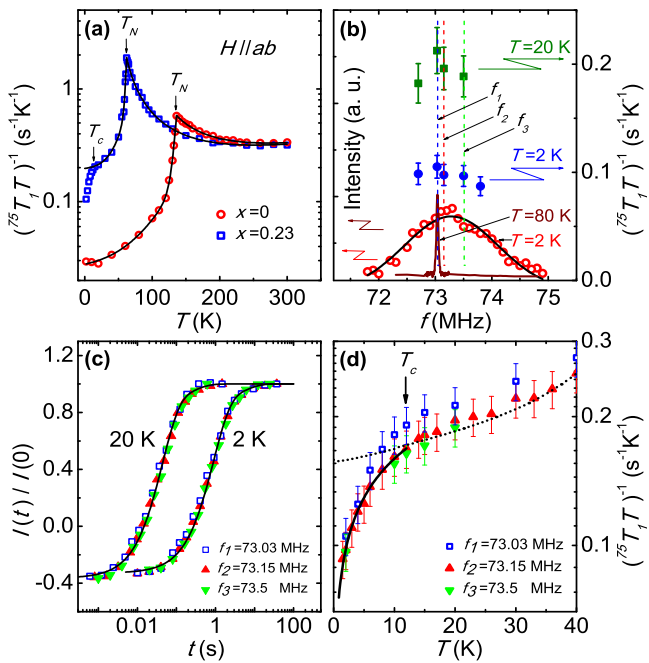}
\caption{\label{slrr4}(color online) (a) $1/^{75}T_1T (T)$ on semilog axes 
for Ba(Fe$_{1-x}$Ru$_{x}$)$_2$As$_2$ ($x$$=$0, 0.23). Solid lines below 
$T_N$ indicate the trend towards a low-temperature constant. (b) $^{75}$As 
spectrum at $T =$ 2 K and 80 K, shown with 1/$^{75}T_1T$ at $T$ = 20 K and 
2 K. (c) $^{75}$As magnetization recovery curve at three frequencies. The 
solid line represents a fit using a single $T_1$ component. (d) $1/^{75}T_1T$ 
at low temperatures for the three frequencies $f_1$, $f_2$, and $f_3$ (see 
text).}  
\end{figure}

Above $T_N$, spin-fluctuation effects are evident in both $^{75}K$ and 
$1/^{75}T_1$. By fitting the relaxation rate to a sum of Curie-Weiss and 
Korringa components, $1/^{75}T_1 = AT/(T - \theta) + BT$, we find that   
$1/^{75}T_1$ in the doped sample has a strong Curie-Weiss contribution 
with $\theta \approx$ 46 K for $H \parallel ab$ [shown as the solid line in 
Fig.~\ref{slrr4}(a)]. Such a form is consistent with low-energy AFM spin 
fluctuations in itinerant systems \cite{Moriya}, and the diverging behavior 
of $1/^{75}T_1$ at $T = T_N$ indicates the magnetic transition. $^{75}K(T)$ 
measures the spin susceptibility at $q = 0$ and therefore shows no 
low-energy AFM spin fluctuations, but its weakly quadratic temperature 
dependence on top of a constant contribution from itinerant electrons is 
fully consistent with local-moment fluctuations \cite{Ma_prb_84}. 
Comparison with BaFe$_2$As$_2$ shows that Ru doping causes no strong 
changes in $^{75}K$ or $1/^{75}T_1$ above $T_N$, as also reported in 
Ref.~\cite{Dey_JPCM_23_475701}. Because these quantities measure magnetic 
correlation effects and the electronic density of states (DOS) on the Fermi 
surface, this weak dependence is consistent with a nearly isovalent doping 
effect \cite{Brouet_PRL_105_087001, Dhaka_prl, Zhang_PRB_79_174530}. 

The superconducting transition is clearly visible in the spin-lattice 
relaxation rate. In Fig.~\ref{slrr4}(a) we show $1/^{75}T_1T (T)$ with $H 
\parallel ab$ for comparison with Fermi-liquid behavior ($1/T_1T$ constant). 
On cooling below 12 K, the spectral weight of the NMR signal is reduced (data 
not shown) and $1/^{75}T_1T$ drops abruptly; both observations are signatures 
of superconductivity fully consistent with the a.c.~susceptibility data. We 
have also measured $1/^{75}T_1$ across the spectrum at $T =$ 2 K and 20 K, 
shown as a function of frequency in Fig.~\ref{slrr4}(b). The reduced, uniform 
value of $1/^{75}T_1T$ at 2 K supports the opening of a uniform superconducting 
gap below $T_C$. We further measured $1/^{75}T_1 (T)$ at selected frequencies 
$f_1$, $f_2$, and $f_3$, taken close to the peak frequency as shown in 
Fig.~\ref{slrr4}(b). Signals at $f_2$ and $f_3$ are clearly due to magnetic 
sites, as their echo intensity is zero above $T_N$. Uniform superconductivity 
is clearly present over a frequency range with at least 70\% of the spectral 
weight, for sites with magnetic moments centered at 0.4 $\mu_B$. Figure 
\ref{slrr4}(c) demonstrates that the $^{75}$As magnetization is fitted very 
well by a single $T_1$ component at all of the selected frequencies, excluding 
any phase separation. In Fig.~\ref{slrr4}(d), $1/^{75}T_1T$ is shown at the 
same three frequencies; that all drop identically at $T_c$ suggests that 
superconductivity in our sample is very homogeneous. 

The abrupt drop in $1/^{75}T_1T$ is unambiguous, local evidence for the 
superconducting gap opening on the magnetic sites. The results detailed in 
the previous paragraph, namely the uniform nature of our $1/T_1$ data and 
the absence of bulk paramagnetism, supplemented by the thermal conductivity 
results of Ref.~\cite{Qiu_PRX_2_011010}, indicate that this superconductivity 
is a homogeneous, bulk phenomenon and that it cannot be a proximity effect 
arising from non-magnetic regions. Our results therefore cannot be interpreted 
as a bulk or microscopic phase separation, and establish clearly a microscopic 
coexistence of SC and AFM in Ba(Fe$_{0.77}$Ru$_{0.23}$)$_2$As$_2$. 

To further understand the Ru doping effect, one may compare $1/^{75}T_1$ 
below $T_N$ for Ba(Fe$_{0.77}$Ru$_{0.23}$)$_2$As$_2$ and BaFe$_2$As$_2$. This
quantity tends to level off to a constant value [Fig.~\ref{slrr4}(a)], as 
expected for a suppression of magnetic fluctuations far below $T_N$. However, 
the constant value is one order of magnitude higher for the doped sample. 
Because $1/^{75}T_1T$ is a measure of the electron DOS on the Fermi surface 
when spin fluctuations are weak, our data show that the density of itinerant 
electrons below $T_N$ is increased by a factor of 3$-$4 due to Ru doping. To 
our knowledge, this clear experimental evidence for doping-induced itinerant 
electrons, going beyond a simple isovalent effect, has not been obtained 
before. These itinerant electrons have significant implications for both 
the AFM and SC phases.

Considering first the AFM phase, the doping-induced itinerant electrons 
may be connected with magnetic dilution. The weakening of antiferromagnetic 
correlations upon Ru doping suggests that the itinerant electrons are either 
non-magnetic or weakly magnetic. The magnetism of the parent compound is 
believed to have a local-moment origin in the valence electrons. Because Ru 
has more extended $d$ orbitals than Fe, the presence of itinerant electrons 
can be understood by shifts of magnetic valence electrons to the Fermi 
surface. Thus our results provide experimental support for the proposed 
magnetic dilution effect \cite{Dhaka_prl, Brouet_PRL_105_087001, Xu_12034699} 
in a band picture. We stress, however, that magnetic dilution may be an 
effect rather than a cause of electron redistribution, and chemical pressure 
is thought to have a key role in Ru doping \cite{Zhang_PRB_79_174530}.

Turning next to the SC phase, the itinerant electrons we observe may be the 
key to a full understanding of superconductivity in the magnetic state. We 
note that a temperature decrease from $T_C$ down to 2 K causes $1/^{75}T_1T$ 
to fall by a factor of two [Fig.~\ref{slrr4}(a)], which suggests that it is 
the itinerant electrons that condense into superconducting pairs. It is highly 
compelling to propose that these itinerant electrons are required for the 
phase coherence of superconductivity, and that a high density of non-magnetic 
or weakly magnetic electrons may be a prerequisite for the coexistence of 
AFM and SC. Our results therefore serve as valuable new input for microscopic 
models of coexistence. 

We stress that we do not observe an appreciable drop of the AFM moment below 
$T_C$ [inset, Fig.~\ref{spec2}(a)], {\it i.e.} we find little evidence of 
competition between the two types of order, in contradiction to some theories 
\cite{Wiesenmayer_PRL_107, Fernandes_PRB_82_014521}. Our data suggest rather 
that the itinerant electrons are non-magnetic, and occupy different parts of 
the Fermi surface from the magnetic electrons. Quantitatively, the ordered 
moment at this doping (0.4 $\mu_B$/Fe) may be too large to be affected 
appreciably by SC, whereas materials with slightly higher doping (lower 
moment) may show some suppression of AFM by SC. 

We close by commenting on the pairing symmetry. The coexistence of AFM 
and SC is consistent \cite{Fernandes_PRB_82_014521} with the $s_{\pm}$ 
superconducting order parameter proposed in other pnictides 
\cite{Nagai_08091197, Ning_PRL_104, Zhang_prb_81, YuW_11011017, 
Baek_PRL, Zheng_11024417, Grafe_PRL_101_047003, Bang_PRB_79_054529}. 
Our results [Fig.~4(d)] show that $1/^{75}T_1T$ is rather large at 2 K, 
even when the relatively high magnetic field is considered; this suggests 
a high density of low-energy excitations in the SC state, which is consistent 
with recent suggestions of nodal superconductivity in underdoped 
Ba(Fe$_{1-x}$Ru$_{x}$)$_2$As$_2$ \cite{Qiu_PRX_2_011010}. ``Accidental'' 
line nodes occurring due to the band dispersion in $k_z$ have been proposed 
in the BaFe$_2$(As$_{0.7}$P$_{0.3}$)$_2$ \cite{Zhang_NP}. Alternatively, the 
coexistence of long-range AFM with SC may in fact cause a nodal gap 
\cite{Maiti_arxiv_1203_0991}. Although we cannot distinguish whether our 
samples fit one or both scenarios, our observation of coexisting AFM and 
SC offers essential input for understanding the excitation spectrum in 
underdoped Fe superconductors.

In summary, we have used $^{75}$As NMR to study high-quality single crystals 
of Ba(Fe$_{0.77}$Ru$_{0.23}$)$_2$As$_2$. The NMR spectra below $T_N$ show that 
the sample is fully magnetic, and are consistent with commensurate 
antiferromagnetism in which the magnetic moments are reduced by comparison 
with BaFe$_2$As$_2$. We find a sharp drop in $1/^{75}T_1T$ below $T_C$ on 
the magnetic sites, which provides unambiguous evidence for the microscopic 
coexistence of bulk superconductivity and antiferromagnetic order. We measure 
a high value of $1/^{75}T_1T$ below $T_N$, which indicates that Ru doping 
induces itinerant electrons, and also a high density of low-energy excitations 
below $T_c$. These observations quantify the effects of Ru doping and provide 
extensive insight into the nature of the coexistence between superconductivity 
and magnetism. 

Work at Renmin University of China was supported by the National Science 
Foundation of China (Grant Nos.~11074304 and 11174365) and the National 
Basic Research Program of China (Nos.~2010CB923004, 2011CBA00112, and 
2012CB921704). Work at POSTECH was supported by the Basic Science Research 
Program (No.~2010-0005669) and by the Max Planck POSTECH/KOREA Research 
Initiative Program (No.~2011-0031558) through the National Research 
Foundation of Korea (NRF). 

{\it Note added.} On completion of this work, we learn of an independent 
NMR study reporting the microscopic coexistence of AFM and SC in the 
underdoped iron superconductor Ba$_{1-x}$K$_x$Fe$_2$As$_2$ \cite{Li_12042434}.


\begin{thebibliography}{36}
\expandafter\ifx\csname natexlab\endcsname\relax\def\natexlab#1{#1}\fi
\expandafter\ifx\csname bibnamefont\endcsname\relax
  \def\bibnamefont#1{#1}\fi
\expandafter\ifx\csname bibfnamefont\endcsname\relax
  \def\bibfnamefont#1{#1}\fi
\expandafter\ifx\csname citenamefont\endcsname\relax
  \def\citenamefont#1{#1}\fi
\expandafter\ifx\csname url\endcsname\relax
  \def\url#1{\texttt{#1}}\fi
\expandafter\ifx\csname urlprefix\endcsname\relax\def\urlprefix{URL }\fi
\providecommand{\bibinfo}[2]{#2}
\providecommand{\eprint}[2][]{\url{#2}}

\bibitem[{\citenamefont{de~la Cruz et~al.}(2008)}]{Dai_Nature_453_899}
\bibinfo{author}{\bibfnamefont{C.}~\bibnamefont{de~la Cruz}}
  \bibnamefont{et~al.}, \bibinfo{journal}{Nature}
  \textbf{\bibinfo{volume}{453}}, \bibinfo{pages}{899} (\bibinfo{year}{2008}).

\bibitem[{\citenamefont{Wiesenmayer et~al.}(2011)}]{Wiesenmayer_PRL_107}
\bibinfo{author}{\bibfnamefont{E.}~\bibnamefont{Wiesenmayer}}
  \bibnamefont{et~al.}, \bibinfo{journal}{Phys. Rev. Lett}
  \textbf{\bibinfo{volume}{107}}, \bibinfo{pages}{237001}
  (\bibinfo{year}{2011}).

\bibitem[{\citenamefont{Pratt et~al.}(2011)}]{Pratt_coexist}
\bibinfo{author}{\bibfnamefont{D.~K.} \bibnamefont{Pratt}}
  \bibnamefont{et~al.}, \bibinfo{journal}{Phys. Rev. Lett.}
  \textbf{\bibinfo{volume}{106}}, \bibinfo{pages}{257001}
  (\bibinfo{year}{2011}).

\bibitem[{\citenamefont{Laplace et~al.}(2009)\citenamefont{Laplace, Bobroff,
  Rullier-Albenque, Colson, and Forget}}]{Laplace_PRB_80}
\bibinfo{author}{\bibfnamefont{Y.}~\bibnamefont{Laplace}} \bibnamefont{et~al.},
  \bibinfo{journal}{Phys. Rev. B} \textbf{\bibinfo{volume}{80}},
  \bibinfo{pages}{140501(R)} (\bibinfo{year}{2009}).

\bibitem{Marsik_PRL_105}
P. Marsik {\it et al.}, Phys. Rev. Lett. {\bf 105}, 057001 (2010).

\bibitem[{\citenamefont{Nandi et~al.}(2010)}]{Nandi_PRL_104}
\bibinfo{author}{\bibfnamefont{S.}~\bibnamefont{Nandi}} \bibnamefont{et~al.},
  \bibinfo{journal}{Phys. Rev. Lett} \textbf{\bibinfo{volume}{104}},
  \bibinfo{pages}{057006} (\bibinfo{year}{2010}).

\bibitem[{\citenamefont{Iye et~al.}(2012)}]{Iye_JPSJ}
\bibinfo{author}{\bibfnamefont{T.}~\bibnamefont{Iye}} \bibnamefont{et~al.},
  \bibinfo{journal}{J. Phys. Soc. Jpn.} \textbf{\bibinfo{volume}{81}},
  \bibinfo{pages}{033701} (\bibinfo{year}{2012}).

\bibitem[{\citenamefont{Drew et~al.}(2009)}]{Drew_NM_8}
\bibinfo{author}{\bibfnamefont{A.~J.} \bibnamefont{Drew}} \bibnamefont{et~al.},
  \bibinfo{journal}{Nature Mater.} \textbf{\bibinfo{volume}{8}},
  \bibinfo{pages}{310} (\bibinfo{year}{2009}).

\bibitem[{\citenamefont{Bao et~al.}(2011)}]{Bao_11020830}
\bibinfo{author}{\bibfnamefont{W.}~\bibnamefont{Bao}} \bibnamefont{et~al.},
  \bibinfo{journal}{Chinese Phys. Lett.} \textbf{\bibinfo{volume}{28}},
  \bibinfo{pages}{086104} (\bibinfo{year}{2011}).

\bibitem[{\citenamefont{Fernandes and
  Schmalian}(2010)}]{Fernandes_PRB_82_014521}
\bibinfo{author}{\bibfnamefont{R.~M.} \bibnamefont{Fernandes}}
  \bibnamefont{and}
  \bibinfo{author}{\bibfnamefont{J.}~\bibnamefont{Schmalian}},
  \bibinfo{journal}{Phys. Rev. B} \textbf{\bibinfo{volume}{82}},
  \bibinfo{pages}{014521} (\bibinfo{year}{2010}).

\bibitem[{\citenamefont{Maiti et~al.}(2012)\citenamefont{Maiti, Fernandes, and
  Chubukov}}]{Maiti_arxiv_1203_0991}
\bibinfo{author}{\bibfnamefont{S.}~\bibnamefont{Maiti}},
  \bibinfo{author}{\bibfnamefont{R.~M.} \bibnamefont{Fernandes}},
  \bibnamefont{and} \bibinfo{author}{\bibfnamefont{A.~V.}
  \bibnamefont{Chubukov}}, \bibinfo{journal}{Phys. Rev. B}
  \textbf{\bibinfo{volume}{85}}, \bibinfo{pages}{144527}
  (\bibinfo{year}{2012}).

\bibitem{Nakai_PRL} 
Y. Nakai {\it et al.}, 
Phys. Rev. Lett. {\bf 105}, 107003 (2010). 

\bibitem[{\citenamefont{Thaler et~al.}(2010)}]{Thaler_PRB_82_014534}
\bibinfo{author}{\bibfnamefont{A.}~\bibnamefont{Thaler}} \bibnamefont{et~al.},
  \bibinfo{journal}{Phys. Rev. B} \textbf{\bibinfo{volume}{82}},
  \bibinfo{pages}{014534} (\bibinfo{year}{2010}).

\bibitem[{\citenamefont{Dhaka et~al.}(2011)}]{Dhaka_prl}
\bibinfo{author}{\bibfnamefont{R.~S.} \bibnamefont{Dhaka}}
  \bibnamefont{et~al.}, \bibinfo{journal}{Phys. Rev. Lett.}
  \textbf{\bibinfo{volume}{107}}, \bibinfo{pages}{267002}
  (\bibinfo{year}{2011}).

\bibitem[{\citenamefont{Brouet et~al.}(2010)}]{Brouet_PRL_105_087001}
\bibinfo{author}{\bibfnamefont{V.}~\bibnamefont{Brouet}} \bibnamefont{et~al.},
  \bibinfo{journal}{Phys. Rev. Lett} \textbf{\bibinfo{volume}{105}},
  \bibinfo{pages}{087001} (\bibinfo{year}{2010}).

\bibitem[{\citenamefont{Xu et~al.}(2012)}]{Xu_12034699}
\bibinfo{author}{\bibfnamefont{N.}~\bibnamefont{Xu}} \bibnamefont{et~al.},
  \bibinfo{journal}{Phys. Rev. B} \textbf{\bibinfo{volume}{86}},
  \bibinfo{pages}{064505} (\bibinfo{year}{2012}).

\bibitem[{\citenamefont{Kim et~al.}(2011)}]{Kim_PRB_83_054514}
\bibinfo{author}{\bibfnamefont{M.~G.} \bibnamefont{Kim}} \bibnamefont{et~al.},
  \bibinfo{journal}{Phys. Rev. B} \textbf{\bibinfo{volume}{83}},
  \bibinfo{pages}{054514} (\bibinfo{year}{2011}).

\bibitem[{\citenamefont{Eom et~al.}(2012)\citenamefont{Eom, Na, Hoch, Kremer,
  and Kim}}]{Kim_JS}
\bibinfo{author}{\bibfnamefont{M.~J.} \bibnamefont{Eom}} \bibnamefont{et~al.},
  \bibinfo{journal}{Phys. Rev. B} \textbf{\bibinfo{volume}{85}},
  \bibinfo{pages}{024536} (\bibinfo{year}{2012}).

\bibitem{rcb}
C. Bernhard {\it et al.}, unpublished (arXiv:1206.7085).

\bibitem[{\citenamefont{Julien et~al.}(2009)}]{Julien_EPL_87_37001}
\bibinfo{author}{\bibfnamefont{M.-H.} \bibnamefont{Julien}}
  \bibnamefont{et~al.}, \bibinfo{journal}{Europhys. Lett.}
  \textbf{\bibinfo{volume}{87}}, \bibinfo{pages}{37001} (\bibinfo{year}{2009}).

\bibitem{ramzea} For a recent review, see A. M. Zhang {\it et al.}, 
Phys. Rev. B, to appear (arXiv:1105.1198). 

\bibitem[{\citenamefont{Zhang and Singh}(2009)}]{Zhang_PRB_79_174530}
\bibinfo{author}{\bibfnamefont{L.}~\bibnamefont{Zhang}} \bibnamefont{and}
  \bibinfo{author}{\bibfnamefont{D.~J.} \bibnamefont{Singh}},
  \bibinfo{journal}{Phys. Rev. B} \textbf{\bibinfo{volume}{79}},
  \bibinfo{pages}{174530} (\bibinfo{year}{2009}).

\bibitem[{\citenamefont{Qiu et~al.}(2012)}]{Qiu_PRX_2_011010}
\bibinfo{author}{\bibfnamefont{X.}~\bibnamefont{Qiu}} \bibnamefont{et~al.},
  \bibinfo{journal}{Phys. Rev. X} \textbf{\bibinfo{volume}{2}},
  \bibinfo{pages}{011010} (\bibinfo{year}{2012}).

\bibitem{Kitagawa_JPSJ}
K. Kitagawa {\it et al.},
J. Phys. Soc. Jpn. {\bf 80}, 033705 (2011). 

\bibitem[{\citenamefont{Kitagawa et~al.}(2008)\citenamefont{Kitagawa, Katayama,
  Ohgushi, Yoshida, and Takigawa}}]{Kita_JPSJ_77_114709}
\bibinfo{author}{\bibfnamefont{K.}~\bibnamefont{Kitagawa}} \bibnamefont{et~al.},
  \bibinfo{journal}{J. Phys. Soc. Jpn.} \textbf{\bibinfo{volume}{77}},
  \bibinfo{pages}{114709} (\bibinfo{year}{2008}).

\bibitem[{\citenamefont{Moriya and Ueda}(1974)}]{Moriya}
\bibinfo{author}{\bibfnamefont{T.}~\bibnamefont{Moriya}} \bibnamefont{and}
  \bibinfo{author}{\bibfnamefont{K.}~\bibnamefont{Ueda}},
  \bibinfo{journal}{Solid State Commun.} \textbf{\bibinfo{volume}{15}},
  \bibinfo{pages}{169} (\bibinfo{year}{1974}).

\bibitem[{\citenamefont{Ma et~al.}(2011)}]{Ma_prb_84}
\bibinfo{author}{\bibfnamefont{L.}~\bibnamefont{Ma}} \bibnamefont{et~al.},
  \bibinfo{journal}{Phys. Rev. B} \textbf{\bibinfo{volume}{84}},
  \bibinfo{pages}{220505(R)} (\bibinfo{year}{2011}).

\bibitem[{\citenamefont{Dey et~al.}(2011)\citenamefont{Dey, Khuntia, Mahajan,
  Sharma, and Bharathi}}]{Dey_JPCM_23_475701}
\bibinfo{author}{\bibfnamefont{T.}~\bibnamefont{Dey}} \bibnamefont{et~al.}, 
  \bibinfo{journal}{J. Phys. Condens. Matter}
  \textbf{\bibinfo{volume}{23}}, \bibinfo{pages}{475701}
  (\bibinfo{year}{2011}).

\bibitem[{\citenamefont{Nagai et~al.}(2008)\citenamefont{Nagai, Hayashi, Nakai,
  Nakamura, Okumura, and Machida}}]{Nagai_08091197}
\bibinfo{author}{\bibfnamefont{Y.}~\bibnamefont{Nagai}} \bibnamefont{et~al.}, 
  \bibinfo{journal}{New J. Phys.} \textbf{\bibinfo{volume}{10}},
  \bibinfo{pages}{103026} (\bibinfo{year}{2008}).

\bibitem[{\citenamefont{Ning et~al.}(2010)}]{Ning_PRL_104}
\bibinfo{author}{\bibfnamefont{F.~L.} \bibnamefont{Ning}} \bibnamefont{et~al.},
  \bibinfo{journal}{Phys. Rev. Lett.} \textbf{\bibinfo{volume}{104}},
  \bibinfo{pages}{037001} (\bibinfo{year}{2010}).

\bibitem[{\citenamefont{Zhang et~al.}(2010)}]{Zhang_prb_81}
\bibinfo{author}{\bibfnamefont{S.~W.} \bibnamefont{Zhang}}
  \bibnamefont{et~al.}, \bibinfo{journal}{Phys. Rev. B}
  \textbf{\bibinfo{volume}{81}}, \bibinfo{pages}{012503}
  (\bibinfo{year}{2010}).

\bibitem[{\citenamefont{Yu et~al.}(2011)}]{YuW_11011017}
\bibinfo{author}{\bibfnamefont{W.}~\bibnamefont{Yu}} \bibnamefont{et~al.},
  \bibinfo{journal}{Phys. Rev. Lett.} \textbf{\bibinfo{volume}{106}},
  \bibinfo{pages}{197001} (\bibinfo{year}{2011}).

\bibitem[{\citenamefont{Baek et~al.}(2009)}]{Baek_PRL}
\bibinfo{author}{\bibfnamefont{S.-H.} \bibnamefont{Baek}} \bibnamefont{et~al.},
  \bibinfo{journal}{Phys. Rev. Lett.} \textbf{\bibinfo{volume}{102}},
  \bibinfo{pages}{227601} (\bibinfo{year}{2009}).

\bibitem[{\citenamefont{Li et~al.}(2011)\citenamefont{Li, Sun, Lin, Su, Hu, and
  Zheng}}]{Zheng_11024417}
\bibinfo{author}{\bibfnamefont{Z.}~\bibnamefont{Li}} \bibnamefont{et~al.},
  \bibinfo{journal}{Phy. Rev. B} \textbf{\bibinfo{volume}{83}},
  \bibinfo{pages}{140506(R)} (\bibinfo{year}{2011}).

\bibitem[{\citenamefont{Grafe et~al.}(2008)}]{Grafe_PRL_101_047003}
\bibinfo{author}{\bibfnamefont{H.-J.} \bibnamefont{Grafe}}
  \bibnamefont{et~al.}, \bibinfo{journal}{Phys. Rev. Lett.}
  \textbf{\bibinfo{volume}{101}}, \bibinfo{pages}{047003}
  (\bibinfo{year}{2008}).

\bibitem[{\citenamefont{Bang et~al.}(2009)\citenamefont{Bang, Choi, and
  Won}}]{Bang_PRB_79_054529}
\bibinfo{author}{\bibfnamefont{Y.}~\bibnamefont{Bang}},
  \bibinfo{author}{\bibfnamefont{H.-Y.} \bibnamefont{Choi}}, \bibnamefont{and}
  \bibinfo{author}{\bibfnamefont{H.}~\bibnamefont{Won}},
  \bibinfo{journal}{Phys. Rev. B} \textbf{\bibinfo{volume}{79}},
  \bibinfo{pages}{054529} (\bibinfo{year}{2009}).

\bibitem[{\citenamefont{Zhang et~al.}(2012)}]{Zhang_NP}
\bibinfo{author}{\bibfnamefont{Y.}~\bibnamefont{Zhang}} \bibnamefont{et~al.},
  \bibinfo{journal}{Nature Phys.} \textbf{\bibinfo{volume}{8}},
  \bibinfo{pages}{371} (\bibinfo{year}{2012}).

\bibitem[{\citenamefont{Li et~al.}(2012)\citenamefont{Li, Zhou, Sun, Yang, Lin,
  and Zheng}}]{Li_12042434}
\bibinfo{author}{\bibfnamefont{Z.}~\bibnamefont{Li}} \bibnamefont{et~al.}, 
  \bibinfo{journal}{unpublished (arXiv:1204.2434).}  

\end{thebibliography}
\end{document}